# Predicting Reactor Antineutrino Emissions Using New Precision Beta Spectroscopy

D. Asner,[1] K. Burns,[1] B. Greenfield,[1] M.S. Kos,[1] J.L. Orrell,[1] M. Schram,[1] B.VanDevender,[1] and D. Wootan[1]

[1]*Pacific Northwest Laboratory, Richland, WA 99352, USA*

(Dated: April 8, 2013)

**ABSTRACT**      A conceptual experimental method for providing a new measurement of the underlying beta decay spectra from fission products is presented. The goal is to provide additional information related to the prediction of the antineutrino emission spectra from fission reactors and assist evaluation of the reactor neutrino anomaly. Submitted as part of the Snowmass 2013 Electronics Proceedings.

Neutrino experiments at nuclear reactors are currently vital to the study of neutrino oscillations. The observed antineutrino rates at reactors are typically lower than model expectations. This observed deficit is called the "reactor neutrino anomaly". A new understanding of neutrino physics may be required to explain this deficit, though model estimation uncertainties may also play a role in the apparent discrepancy. PNNL is currently investigating an experimental technique that promises reduced uncertainties for measured data to support these hypotheses and interpret reactor antineutrino measurements. The experimental approach is to 1) direct a proton accelerator beam on a metal target to produce a source of neutrons, 2) use spectral tailoring to modify the neutron spectrum to closely simulate the energy distribution of a power reactor neutron spectrum, 3) irradiate isotopic fission foils ($^{235}$U, $^{238}$U, $^{239}$Pu, $^{241}$Pu) in this neutron spectrum so that fissions occur at energies representative of a reactor, 4) transport the beta particles released by the fission products in the foils to a beta spectrometer, 5) measure the beta energy spectrum, and 6) invert the measured beta energy spectrum to an antineutrino energy spectrum.

A similar technique using a beta spectrometer and isotopic fission foils was pioneered in the 1980's at the ILL thermal reactor. Those measurements have been the basis for interpreting all subsequent antineutrino measurements at reactors. A basic constraint in efforts to reduce uncertainties in predicting the antineutrino emission from reactor cores is any underlying limitation of the original measurements. This may include beta spectrum energy resolution, the absolute normalization of beta emission to number of fission, statistical counting uncertainties, lack of $^{238}$U data, the purely thermal nature of the ILL reactor neutrons used, etc. An accelerator-based neutron source that can be tailored to match various reactor neutron spectra provides an advantage for control in studying how changes in the neutron spectra (i.e. "in the reactor core") affects the resulting fission product beta spectrum. Furthermore, the $^{238}$U antineutrino spectrum, which has not been measured, can be studied directly because of the enhanced 1 MeV fast neutron flux available at the accelerator source. A facility such as the Project X Injector Experiment (PXIE) 30 MeV proton linear accelerator at Fermilab is being considered for this experiment. The hypothesis is that a new approach utilizing the flexibility of an accelerator neutron source with spectral tailoring coupled with a careful design of an isotopic fission target and beta spectrometer and the inversion of the beta spectrum to the neutrino spectrum will allow further reduction in the uncertainties associated with prediction of the reactor antineutrino spectrum.

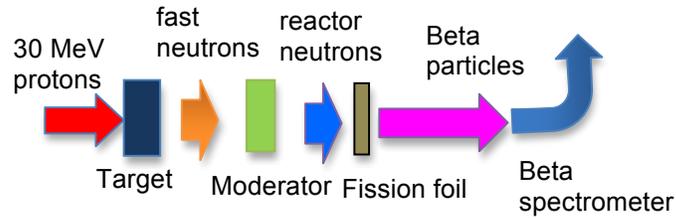

Figure 1.  Experimental Approach

1. **Neutron spectrum tailoring to mimic neutron spectrum in reactor**.  A proton accelerator will be used to produce neutrons from a cooled metal target.  A major advantage of an accelerator neutron source over a neutron beam from a thermal reactor is that the fast neutrons can be slowed down or tailored to approximate various power reactor spectra, thereby allowing investigation of neutron spectral sensitivities that can contribute to reducing the overall uncertainty in the neutrino spectrum. Conversely the neutron spectral tailoring can be matched to the thermal ILL spectrum or left as a predominately fast neutron spectrum for comparative studies. Such neutron spectral tailoring for measurement sake is not available at a reactor.

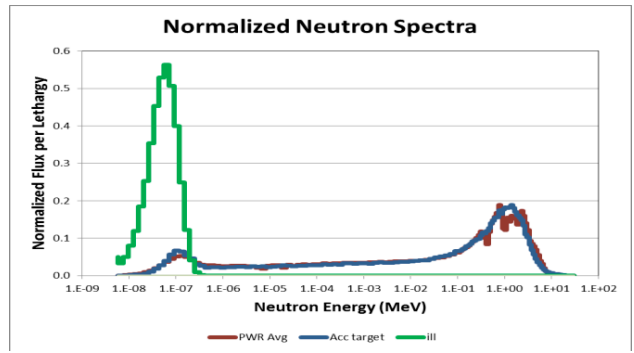

Figure 2.  Spectral Tailoring can Reproduce Reactor Spectrum from Accelerator Source

2. **Fission foil target.** Fission foil targets containing a small (mg) isotopic fission target of the primary reactor isotopes $^{235}$U, $^{238}$U, $^{239}$Pu, or $^{241}$Pu will be used. These foil targets will be optimized for the neutron field and contain the fission products, but allow the delayed fission betas to be released so they can be collected in the beta spectrometer. Attention will be given to measuring secondary constraints on the absolute fission rate such as gamma-rays from specific fission products after irradiation, potentially even during irradiation, and using additional calibrated fission foils designed solely to provide a check on the neutron flux and spectrum produced by the spectral tailoring.

3. **Precision beta spectra measurements from fission products.**  The data underlying all reactor antineutrino flux calculations were collected in the 1980s at ILL.  A magnetic spectrometer (BIL) analyzed the energy of beta decay electrons with resolution 50—100 keV in the energy range 1—10 MeV.  The measurements were very well made.  The primary uncertainty in the data is the precision with which the energy dependent detection efficiency is known.  We are exploring independent beta spectroscopy methods with a focus on understanding of detection efficiencies and other systematic uncertainties.

4. **Deconvolve neutrino spectra using measured beta spectra.** Recent reevaluations of the fission neutrino spectrum based on 1980's reactor beta spectrum measurement data indicate a +3% shift in the neutrino spectrum with a 3-10% uncertainty. Deconvolution of the measured beta spectra into an antineutrino spectrum is necessary, as is ensuring an understanding of matching a given fission product beta spectrum to a given reactor core. We intend to document specific examples of how to map the results of the results of this beta spectrum measurement to operating reactor cores.

The major challenges are to identify and control sources of experimental uncertainties in the measured data and methodology, carefully propagate the correlated and uncorrelated uncertainties through the analysis, and develop an experimental system that will fit within the physical and operational constraints of the irradiation facility while providing highly accurate fission beta energy spectra data. Ultimately the goal is to reduce the experimental uncertainty associated with prediction of reactor antineutrino spectra.